\documentclass{sig-alternate-05-2015}
\usepackage[T1]{fontenc}
\usepackage[utf8]{inputenc}
\usepackage{mathptmx}

\begin{document}

% Copyright
%\setcopyright{acmcopyright}
%\setcopyright{acmlicensed}
%\setcopyright{rightsretained}
%\setcopyright{usgov}
%\setcopyright{usgovmixed}
%\setcopyright{cagov}
%\setcopyright{cagovmixed}
\toappear{} 

% DOI
%\doi{}

% ISBN
%\isbn{}

%Conference
% \conferenceinfo{XXXXXX}{XXXXXXX}

% \acmPrice{\$15.00}

%
% --- Author Metadata here ---
% \conferenceinfo{WOODSTOCK}{'97 El Paso, Texas USA}
%\CopyrightYear{2007} % Allows default copyright year (20XX) to be over-ridden - IF NEED BE.
%\crdata{0-12345-67-8/90/01}  % Allows default copyright data (0-89791-88-6/97/05) to be over-ridden - IF NEED BE.
% --- End of Author Metadata ---

\title{Your Activities of Daily Living (YADL):  An Image-based Survey Technique for Patients with Arthritis}
%
% You need the command \numberofauthors to handle the 'placement
% and alignment' of the authors beneath the title.
%
% For aesthetic reasons, we recommend 'three authors at a time'
% i.e. three 'name/affiliation blocks' be placed beneath the title.
%
% NOTE: You are NOT restricted in how many 'rows' of
% "name/affiliations" may appear. We just ask that you restrict
% the number of 'columns' to three.
%
% Because of the available 'opening page real-estate'
% we ask you to refrain from putting more than six authors
% (two rows with three columns) beneath the article title.
% More than six makes the first-page appear very cluttered indeed.
%
% Use the \alignauthor commands to handle the names
% and affiliations for an 'aesthetic maximum' of six authors.
% Add names, affiliations, addresses for
% the seventh etc. author(s) as the argument for the
% \additionalauthors command.
% These 'additional authors' will be output/set for you
% without further effort on your part as the last section in
% the body of your article BEFORE References or any Appendices.

% \numberofauthors{8} %  in this sample file, there are a *total*
% % of EIGHT authors. SIX appear on the 'first-page' (for formatting
% % reasons) and the remaining two appear in the \additionalauthors section.
% %

\vspace{-3mm}

\author{
Longqi Yang\textsuperscript{\textdagger,$\ddagger$}, Diana Freed\textsuperscript{$\ddagger$}, Alex Wu\textsuperscript{*}, Judy Wu\textsuperscript{$\ddagger$}, John P. Pollak\textsuperscript{$\ddagger$}, Deborah Estrin\textsuperscript{\textdagger,$\ddagger$}\\
\affaddr{\textsuperscript{\textdagger}Department of Computer Science, Cornell University; \textsuperscript{$\ddagger$}Cornell Tech} \\
\affaddr{\textsuperscript{*}MFA Design for Social Innovation, School of Visual Art}\\
\email{ylongqi@cs.cornell.edu; \{diana.freed, alexmtwu, judyzhaoxinwu\}@gmail.com;
}\\
\email{\{jpp9, de226\}@cornell.edu}
}

\vspace{-3mm}

\maketitle

\begin{abstract}
Healthcare professionals use Activities of Daily Living (ADL) to characterize a patient's functional status and to evaluate the effectiveness of treatment plans. ADLs are traditionally measured using standardized text-based questionnaires and the only form of personalization is in the form of question branching logic. Pervasive smartphone adoption makes it feasible to consider more frequent patient-reporting on ADLs. However, asking generic sets of questions repeatedly introduces user burden and fatigue that threatens to interfere with their utility. We introduce an approach called \textbf{YADL} (\textbf{Y}our \textbf{A}ctivities of \textbf{D}aily \textbf{L}iving) which uses images of ADLs and personalization to improve survey efficiency and the patient-experience. It offers several potential benefits: wider coverage of ADLs, improved engagement, and accurate capture of individual health situations. In this paper, we discuss our system design and the wide applicability of the design process for survey tools in healthcare and beyond. Interactions with  with a small number of patients with Arthritis throughout the design process have been promising and we share  detailed insights.
\end{abstract}
\vspace{-3mm}
%A category including the fourth, optional field follows...

% We no longer use \terms command
%\terms{Theory}
\vspace{-3mm}
\vspace{-3mm}
\section{Introduction}
Measuring and understanding an individual's functional state is key to effective treatment of patients with chronic conditions\cite{matsumura1983determination, stewart1989functional}. In this paper, we focus on Patients with Arthritis (PwA), including Rheumatoid Arthritis (RA) and Osteoarthritis (OA). The symptoms and health conditions of PwA are largely manifested in their performances of Activities of Daily Living (ADL) \cite{self1969assessment}, due to muscle weakness, inflammation and limitations caused by joint pain and stiffness \cite{archenholtz2008validity, katz1995impact}. Timely active recordings of the difficulties in performing ADLs by patients informs clinicians of treatment effectiveness and clinical changes \cite{katz1983assessing}. Careful analysis of ADL functional trends is also helpful in prioritizing and scheduling rehabilitative intervention \cite{legg2006occupational}. As the most widely adopted method of collecting information with respect to ADL functionality \cite{katz1983assessing}, a variety of short and long-form questionnaires have been designed and evaluated in both inpatient and outpatient settings \cite{ciro2015instrumental, promis2012patient}. While the answers to questions have been validated as correlated and effective to varying degrees \cite{archenholtz2008validity}, our formative interviews identified three inherent limitations to these static forms:

\textbf{Coverage:} Although activities included in a single standardized questionnaire have been carefully selected to be clinically relevant, it is not possible for these to capture all ADLs that might be difficult for all patients. For instance, in the Boston AM-PAC form \cite{haley2004short}, \textit{walking}, \textit{running} and \textit{bending} are included as they are common and basic measures of mobility, but there are no questions pertaining to \textit{turning a doorknob} or \textit{typing on a keyboard}. For many patients, these omissions would result in failure to capture challenges related to joint stiffness and an overall under-representation of ADL difficulty. This concern arose in our patient interviews and will be discussed in the following sections. 

\textbf{Legibility:} Some items presented in questionnaires are designed specifically to capture patients' difficulties with fine-grained ADLs but may not be relevant to an individual patient. For instance, the following activity descriptions in the standard Boston AM-PAC form, ``\textit{Bending over from a standing position to pick up a piece of clothing from the floor without holding onto anything}'' and ``\textit{Standing up from a low, soft couch}'', are sometimes too detailed or not relevant to the individual patient experiences. This impacts both the  user experience and effectiveness of the survey.

\textbf{Availability:} Standard ADL assessments are typically only completed before, during or after a clinical encounter in an office setting. For example, the Health Assessment Questionnaire (HAQ) \cite{bruce2003stanford} is administered when the patient registers for a clinical visit \cite{pincus2004quantitative}. As a result, the patient's day to day ADLs and variability in performance are usually undersampled \cite{mancuso1995does}. Additionally, data is subject to considerable bias as answers are most often based on patients' retrospective recall \cite{stone2002capturing}.

We present our system named \textbf{YADL} (\textbf{Y}our \textbf{A}ctivities of \textbf{D}aily \textbf{L}iving) - an image-based survey inspired by Photographical Affect Meter (PAM) \cite{pollak2011pam}, designed to improve both fidelity and user experience in reporting ADLs. Compared with traditional questionnaires, YADL is designed to be personalized, engaging and easily accessible. (1) \textit{Personalized experiences:} Instead of comprehensively listing all possible ADLs via word-based descriptions, we leverage images' inherent ambiguities that allow for personalized interpretation and thus capture fine-grained ADL experiences; moreover, the inventory of images can be further tailored to specific demographics and disease contexts. (2) \textit{Patient engagement:} We use representative activity images that are quick and easy to interpret. We argue that these photos are easier to understand and more effective at capturing ADL's data as well. (3) \textit{Wide availability:} YADL can be accessed via a web browser on desktop computers or on mobile devices so that patients can complete YADL at their convenience to keep timely records of their functional states. Photos also provide greater accessibility for lower literacy and non-english-speaking patients.

We have designed YADL using a patient-driven participatory design process to iteratively explore and obtain feedback from clinicians and arthritis patients. Preliminary user testing suggests that patients' responses on YADL exhibit consistency with two widely-used instruments, i.e. Boston AM-PAC \cite{haley2004short} and WOMAC \cite{mcconnell2001western}. In addition, the feedback collected during the design process uncovers several added benefits from YADL that could provide unprecedented information in understanding patients' health conditions. 

In this paper, we present the detailed design approach and discuss the lessons learned throughout the process of user interviews, rapid prototyping, user testing and iterative refinement. We further propose that this exploration of image-based applications as a replacement for short and long-form text questionnaires will inspire new work in healthcare and beyond. Finally, the iterative design process that we adopt in this paper can be widely applied to the general task of standard survey improvements and the transition from text-dominated interfaces to visually-rich interfaces. Future work, informed by the detailed feedback in the preliminary user testing, will be conducted to validate these results with a larger randomized sample.
\vspace{-3mm}
\section{The System}

YADL is a responsive web application designed to be accessible across a wide range of devices. At the beginning, the patient is presented with an activity image and asked to indicate ``\textit{How difficult is this activity for you on a difficult day?}'' The patient then selects his/her response from the choices of Easy, Moderate and Hard, or Skip if the image is irrelevant or hard to interpret. After making a selection, another image is immediately presented. This continues until all images in the database have been evaluated. An example of user interface is shown in Fig.1(a). Patients with arthritis may have accessibility issues due to the stiffness and/or pain in the fingers. Our design addresses these concerns by using color-coded large buttons and a simple one tap gesture.

\begin{figure}
\centering
  \includegraphics[width=0.7\columnwidth]{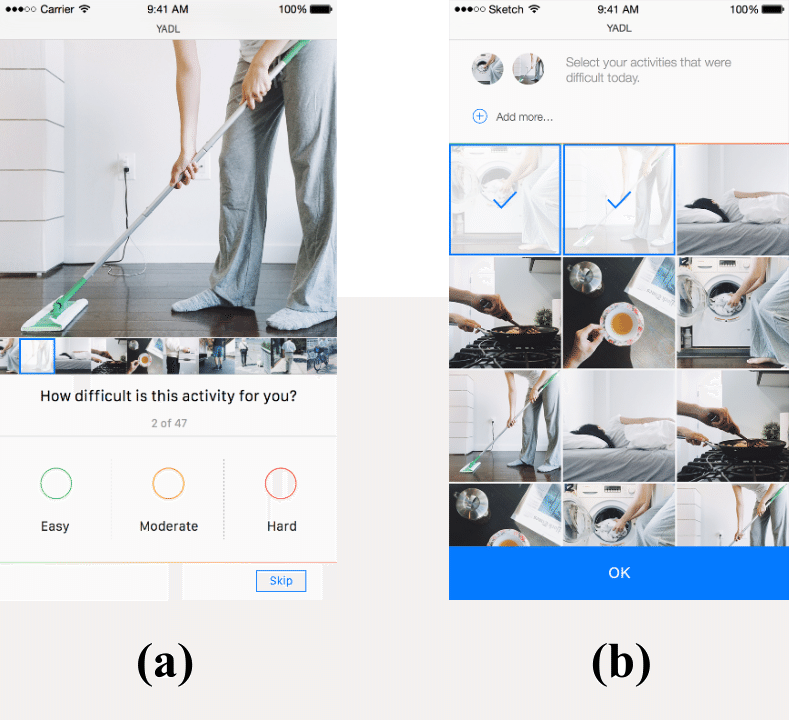}
  \caption{YADL user interface on mobile web browser}
\label{fig:yadl}
\vspace{-4mm}
\end{figure}

\vspace{-3mm}
\section{Design Process}
\vspace{-1mm}
\subsection{Design opportunities identification}
To identify and explore the design opportunities, we conducted both telephone and in person interviews with three arthritis patients to understand their previous experiences with traditional ADL questionnaires. 

\textbf{Legibility.} When we asked the question ``\textit{Did you feel that the traditional questionnaire captured your health functionality?}'' in the interview, two patients responded, ``\textit{No... I rush through paper forms at MD…}'' (P1) and ``\textit{No, I go through as fast as I can...}'' (P2), suggesting that the form-based approach to ADLs is ineffective in engaging patients and insufficient for capturing functional states of ADLs. Significant opportunity exists to improve on the patient experience and hence reporting fidelity.

\textbf{Availability.} The granularity of standard assessments is greatly affected by the frequency of clinician encounters, We asked the patients ``\textit{How often do you see your doctor or another healthcare professional for your RA/OA/PA?}''. Participants maintained office visits at  ``\textit{6 month}'' (P1), ``\textit{once a month}'' (P2) and ``\textit{A couple of times a year}'' (P3). As a result, patients' everyday or even week to week variations of ADL functionality are not captured and communicated in the current system. Retrospective recall, at the time of the clinical encounter is subject to recall bias. With the mobile first principle in mind, the survey can be easily accessed and completed via mobile devices and provide opportunities for fine-grain data capture.

\textbf{Coverage.} When we asked the patients to describe how pain and flare-ups impact their related daily life, responses included, ``\textit{I cannot twist a cap and open certain doors... I am afraid of getting hit by a bike so I take a long walk everyday and only walk in the same neighborhood}'' (P3), ``\textit{Stepping into a bathtub is dangerous}'' (P2). The activities mentioned in the responses, e.g. \textit{open doorknobs}, \textit{twist a cap} and \textit{step into bathtub}, are not reflected in traditional questionnaires, such as Boston AM-PAC and WOMAC, so that they are insufficient to capture patients' actual health conditions. Therefore, survey instruments that cover wider range of ADLs are needed to alleviate this limitation.

\subsection{Key activity identification}
\begin{table}[]
\centering

\label{my-label}
\begin{tabular}{|p{8cm}|}
\hline
Bathing, Feeding, Dressing independently, Walking, Climbing stairs, Rising from chair, Yard work, Driving, Errands, Cooking, Feeds self, Holds drinking cup, Washing, Lifting groceries, Hair brushing, Buttoning shirt, Zipping coat, Exercise, Work, Cooking, Kneeling, Bending , Stooping, Gardening, Shaving, Brushing teeth, Open doors, Faucet handles, Door knobs, Hold pens, Hold silverware, Pick up glass, Open jar, Mopping, Sweeping, Vacuuming, Cutting food, Squeezing toothpaste, Blowdrying hair, Open carton of milk, Socializing, Using scissors, Parties, Theatre, Dining out \\ \hline
\end{tabular}
\caption{Identified activity list}
\vspace{-3mm}
\label{tbl:list}
\end{table}

Activity images are the key component of our application. We started by identifying key activities through a two-part process. First, we performed a comprehensive literature search to determine the most commonly used ADL assessment forms and extracted the activity inventory from each. We spoke with rheumatologists and occupational therapists and conducted a retrospective review of ADL assessments as applied to the arthritis population. The activity phrases were extracted from the main verbs in the articles, e.g. \textit{standing up}, \textit{walking}, \textit{running} etc., such that they are general enough for patients to have personalized interpretations. Second, we identified online forums and communities as important venues for understanding real situations of patients and the hard ADLs that are common among the population. We gathered data from  support and wellness communities, e.g. \textit{inspire}; community forums on patient advocacy organizations, e.g. \textit{arthritis.org}; social networks, e.g. \textit{MyRATeam}; peer-to-peer support forums, e.g. \textit{patientslikeme} and \textit{rheum4us}; and news websites, e.g. \textit{medium.com}, where patients regularly shared their health conditions and day-to-day personal stories. From their descriptions, we extracted activity phrases pertaining to everyday ADLs, i.e. common verbs that were mentioned in the online posts, and use them as additions to traditional questionnaires. With an aggregated list of ADLs based on common assessments and online posts, we then worked with clinicians and patients to review and further refined the list. The final list of identified key items are presented in Table.\ref{tbl:list}.

\subsection{Image Matching and iterative refinement}

For each activity phrase, we searched on Flickr and Google for images with Creative Commons Licenses for our initial images which we reviewed with a clinician and 3 patients. However, through the interviews, we found that stock images alone did not include appropriate \textbf{details} and enough \textbf{demographic diversity}. For example, we presented patients with images of a lever doorknob and learned that difficulty is consistent with round doorknobs (P1, P2). Similarly, we showed individuals walking on a paved surface and through user interviews, we learned the importance of allowing patients to customize images to represent gravel, snow, grass, dirt and sand as different surfaces were relevant for different individual patients (P1, P2). 

To complement the limitations of the public images, we then worked with a photographer and recruited volunteers to pose for photo shoots in various environmental settings, e.g. walking on different surfaces. All of the images were cropped to focus on the particular activities and none of them include the subjects' faces. We further improved the image library by getting feedback on several variations from both clinicians and patients. Each matched photo was shot several times, in different settings, to assure that the images we captured and tested can be understood quickly. As a result, 47 activity images were filtered for YADL. A grid of sample images is shown in Fig.2. Through the patient interviews, we further discovered that the images included possess two added benefits in understanding patients' health conditions, which are discussed in detail next: 

\begin{figure}
\centering
  \includegraphics[width=1.0\columnwidth]{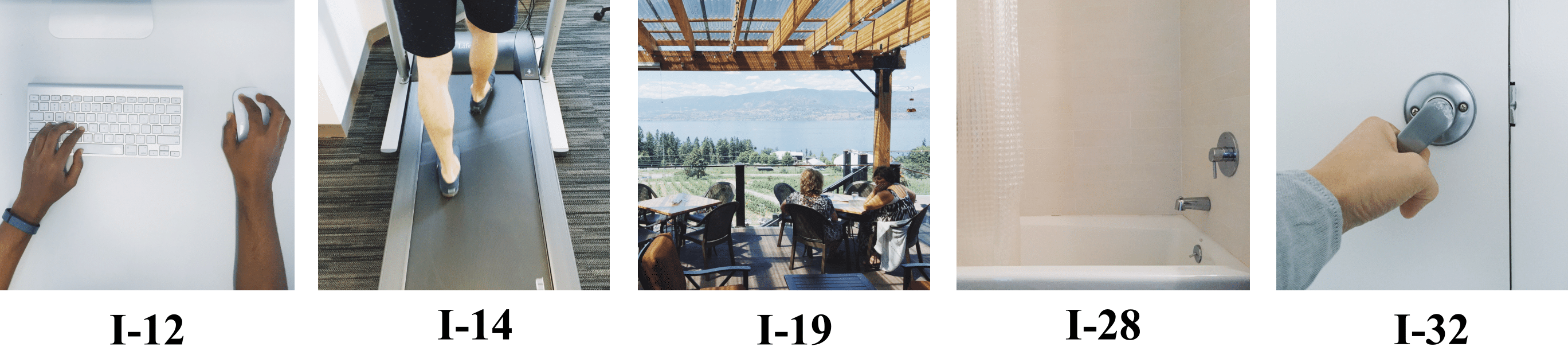}
  \caption{Sample activity images used in YADL (Indexes below)
}
\label{fig:sample}
\vspace{-3mm}
\end{figure}

\textbf{Encourage personal interpretations:} Specific images are inherently ambiguous for some individuals due to cultural and characteristic differences. This phenomenon is shown in the data that we collected. However, we observe that this type of image also provided opportunities for personal interpretations and thus helped to explore hard ADLs that can not be covered by simply listing them with text descriptions. For example, regarding image I-19, although P3 felt unclear about the content, ``\textit{Watching the horizon? Not clear…}'', P1 and P2 successfully interpreted it and P2 even brought up her related personal story. ``\textit{Socializing}'' (P1), ``\textit{Sitting in a restaurant...I can't do it with neck and back pain.}'' (P2). This also happened for image I-28. While P3 felt it to be unclear,  ``\textit{Picture is not clear. Are you asking if I turn on the water?...}'' , P2 made strong statement of how this activity will affect her life, saying ``\textit{This is a bathtub...I can't bend my body to get into a bath.}'' These personalized experiences are hard to capture with static questionnaires and we believe that visual contents could play an important role in engaging and understanding PwA.

\textbf{Catalyze story sharing:} The activity images could also generate conversations between the patients and the their care providers about their pain-related personal stories, which are helpful in understanding the changes in their health conditions and impact on their day-to-day lives. For example, the image of people socializing (I-19) generated a discussion about the impact of RA on social life and how ``\textit{flare-ups force me to stay home and made me depressed.}''(P2) Images of door handles (I-32) reminded two patients (P1, P2) of specific use cases of lever doorknobs vs. twist doorknobs, which highlights individual patient concerns about recent changes regarding management of daily care. Typically, within limited clinical office visit time it is difficult to hit upon the right questions for each patient to evaluate functionality and go through detailed dialogs, the illustrative experiences that brought by images could be an useful addition for physician assistants, nurse practitioners, occupational and physical therapists, care coordinators and family members to understand patients' individual situations.
\vspace{-3mm}
\section{Preliminary user testing}

Our design process included iterative testing with three patients diagnosed with different types of arthritis (P1:OA, P2:RA, P3:OA). Participants were recruited from Facebook patient communities. Testing consisted of three components: (1) Each participant was asked to assess him/herself using all 47 images in our exploratory inventory for YADL and his/her responses were recorded. (2) Each patient was granted access to a web-based version of the Boston AM-PAC and WOMAC forms and instructed to fill out both forms respectively. (3) A telephone interview followed during which the participants rated the experience and gave feedback. By comparing their choices on different types of instruments and analyzing their conversations with the interviewers, we were able to compare the functionality of the YADL survey to that of standard instruments. The feedback provided by the participants suggests that YADL could be an effective and promising tool for capturing the fine-grained functional state of ADLs for patients with arthritis.

\textbf{Comparison of YADL to standard assessments.} We compared the results of YADL with those from gold-standard ADL surveys and our preliminary results suggest that YADL can be easily understood by the patients and that their selections accurately reflect the functional states of ADLs. In our testing, we chose the Boston AM-PAC form and WOMAC as the assessments. To make our system and text-based questionnaires directly comparable, we quantified the available choices in YADL, i.e. Easy = 0, Moderate = 1, Hard = 2. Also, each question in the standard form was pre-matched with a set of related activity images (i.e. sharing the same verb). As a result, 18/47 and 25/47 images were necessary and sufficient to match all questions in the AM-PAC form and WOMAC respectively. The metrics that we compare are the highest response value on the image set and the patient's choice on the questionnaire. While the number of patients is too small for real inference, \textit{Pearson correlations} are 0.743 and 0.747 between patient responses to YADL and AM-PAC and WOMAC suggesting that the photographical interface on YADL can be effective and accurate in capturing difficulties of ADLs for PwA.

\textbf{Covering a wider range of ADLs.} Compared with standard assessments such as Boston AM-PAC and WOMAC, YADL captures wider range of ADLs that are difficult for the patients. In the question-image matching process mentioned above, 29/47 and 22/47 images remained uncovered by AM-PAC and WOMAC respectively. Among all these unmatched activity photos, 15/29 and 11/22 of them are expressed as hard ADLs by at least one of the participants. For instance, P2 clicked hard for image I-12 (Fig. 2), \textit{typing on keyboard}, image I-19, \textit{sitting in the restaurant} (\textit{socializing}), and image I-14, \textit{walking on the treadmill}, which are missing in both forms. The above results justify the value of the comprehensive literature search and the activity phrases extraction incorporated into YADL design as this process ensures wider coverage of ADLs with our system.

\textbf{Overall patient evaluation.} In our testing among three patients, YADL was favored by two of the participants over standard questionnaires. For the question ``\textit{Which method would do you think best captures how you are doing on a day to day basis? (Questionnaire/Neither/YADL)}'', P1 and P2 expressed strong preferences for YADL. ``\textit{... liked pictures with people in them...}'' (P1), ``\textit{Liked image much more ... somewhat I like the one with pictures,  just easier to see it and feel it.}'' (P2). Although P3 responded neutrally to this question, ``\textit{either could work. I am used to the forms and not used to the pictures but they are fine.}'', we did observe that YADL engaged her with finer-grained ADL captured.
\vspace{-1mm}
\section{Discussion}

\subsection{Generalization of design process}

The iterative process that we present in this paper is applicable to other domains of human computer interaction where current text-dominated surveys and interfaces would benefit from this visually-rich approach:

\textbf{Key item identification from broad assessments and online articles.} With the overwhelming amount of information provided by variable standard questionnaires, the process of aggregating and harvesting key items would benefit the design in terms of wide coverage in the initial prototype. In addition, online articles are important sources of information about individual situations. Incorporating this information could potentially provide unprecedented utilities for end users. This process could also be used to refine existing forms as the information from patients' daily posts could provide directions for the changes.

\textbf{Fast iterative refinement with inputs from field professions.} It's critical to carry the iterative design fast along with field professions. This is the process in which survey items are adjusted and refined quickly based on the inputs from domain experts. In the design of text-to-visual interfaces, domain knowledge is needed for the conversion so as to guarantee that the alternative representations are understandable and scientifically valid.

\subsection{Personalized self-report tool}

Personalized activity images that patients indicate as hard in the survey can further fuel a customizable self-report tool that enables timely and quick-tap records of their functional states, as Fig. 1(b) shows. The tool could base on the reduced custom set of images that patient has chosen and thus provide finer-grained data points for clinicians to understand patients’ health conditions.
\vspace{-3mm}
\section{Conclusion and Future Work}

In this paper, we introduced the design of YADL, an image-based survey technique for patients with arthritis. The iterative design process presented in this paper: \textbf{identify key items from broader resources} and \textbf{regularly elicit inputs from field professions}, can be adapted to transform text-dominated interfaces to visually-rich interfaces more generally in healthcare and beyond. Next steps for YADL will be the validation of the system at scale and development the self report tool. We will scale the number of participants and test it in a longitudinal study to validate its value in understanding ADL and patient function.
%
% The following two commands are all you need in the
% initial runs of your .tex file to
% produce the bibliography for the citations in your paper.

\bibliographystyle{abbrv}

\bibliography{sigproc}  % sigproc.bib is the name of the Bibliography in this case

\begin{thebibliography}{10}

\bibitem{archenholtz2008validity}
B.~Archenholtz and B.~Dellhag.
\newblock Validity and reliability of the instrument performance and
  satisfaction in activities of daily living (ps-adl) and its clinical
  applicability to adults with rheumatoid arthritis.
\newblock {\em Scandinavian Journal of Occupational Therapy}, 15(1):13--22,
  2008.

\bibitem{bruce2003stanford}
B.~Bruce and J.~F. Fries.
\newblock The stanford health assessment questionnaire: a review of its
  history, issues, progress, and documentation.
\newblock {\em The Journal of Rheumatology}, 30(1):167--178, 2003.

\bibitem{ciro2015instrumental}
C.~A. Ciro, M.~P. Anderson, L.~A. Hershey, C.~I. Prodan, and M.~B. Holm.
\newblock Instrumental activities of daily living performance and role
  satisfaction in people with and without mild cognitive impairment: A pilot
  project.
\newblock {\em American Journal of Occupational Therapy},
  69(3):6903270020p1--6903270020p10, 2015.

\bibitem{haley2004short}
S.~M. Haley, P.~L. Andres, W.~J. Coster, M.~Kosinski, P.~Ni, and A.~M. Jette.
\newblock Short-form activity measure for post-acute care.
\newblock {\em Archives of Physical Medicine and Rehabilitation},
  85(4):649--660, 2004.

\bibitem{katz1995impact}
P.~P. Katz.
\newblock The impact of rheumatoid arthritis on life activities.
\newblock {\em Arthritis \& Rheumatism}, 8(4):272--278, 1995.

\bibitem{katz1983assessing}
S.~Katz.
\newblock Assessing self-maintenance: activities of daily living, mobility, and
  instrumental activities of daily living.
\newblock {\em Journal of the American Geriatrics Society}, 31(12):721--727,
  1983.

\bibitem{self1969assessment}
B.~E.~M. Lawton M.~Powell.
\newblock Assessment of older people: self-maintaining and instrumental
  activities of daily living.
\newblock 1969.

\bibitem{legg2006occupational}
L.~Legg, A.~Drummond, and P.~Langhorne.
\newblock Occupational therapy for patients with problems in activities of
  daily living after stroke.
\newblock {\em The Cochrane Library}, 2006.

\bibitem{mancuso1995does}
C.~A. Mancuso and M.~E. Charlson.
\newblock Does recollection error threaten the validity of cross-sectional
  studies of effectiveness?
\newblock {\em Medical Care}, pages AS77--AS88, 1995.

\bibitem{matsumura1983determination}
N.~Matsumura, H.~Nishijima, S.~Kojima, F.~Hashimoto, M.~Minami, and H.~Yasuda.
\newblock Determination of anaerobic threshold for assessment of functional
  state in patients with chronic heart failure.
\newblock {\em Circulation}, 68(2):360--367, 1983.

\bibitem{mcconnell2001western}
S.~McConnell, P.~Kolopack, and A.~M. Davis.
\newblock The western ontario and mcmaster universities osteoarthritis index
  (womac): a review of its utility and measurement properties.
\newblock {\em Arthritis care \& research}, 45(5):453--461, 2001.

\bibitem{pincus2004quantitative}
T.~Pincus, T.~Sokka, and A.~Kavanaugh.
\newblock Quantitative documentation of benefit/risk of new therapies for
  rheumatoid arthritis: patient questionnaires as an optimal measure in
  standard care.
\newblock {\em Clin Exp Rheumatol}, 22(5 Suppl 35):S26--33, 2004.

\bibitem{pollak2011pam}
J.~P. Pollak, P.~Adams, and G.~Gay.
\newblock Pam: a photographic affect meter for frequent, in situ measurement of
  affect.
\newblock In {\em Proceedings of the SIGCHI Conference on Human Factors in
  Computing Systems}, pages 725--734. ACM, 2011.

\bibitem{promis2012patient}
P.~PROMIS.
\newblock Patient reported outcomes measurement information system.
\newblock {\em National Health Institute}, 2012.

\bibitem{stewart1989functional}
A.~L. Stewart, S.~Greenfield, R.~D. Hays, K.~Wells, W.~H. Rogers, S.~D. Berry,
  E.~A. McGlynn, and J.~E. Ware.
\newblock Functional status and well-being of patients with chronic conditions:
  results from the medical outcomes study.
\newblock {\em Jama}, 262(7):907--913, 1989.

\bibitem{stone2002capturing}
A.~A. Stone and S.~Shiffman.
\newblock Capturing momentary, self-report data: A proposal for reporting
  guidelines.
\newblock {\em Annals of Behavioral Medicine}, 24(3):236--243, 2002.

\end{thebibliography}
% You must have a proper ".bib" file
%  and remember to run:
% latex bibtex latex latex
% to resolve all references
%
% ACM needs 'a single self-contained file'!
%
%APPENDICES are optional
%\balancecolumns

\end{document}